\documentclass[aps,groupedaddress]{revtex4}

\usepackage{epsfig}
\usepackage{latexsym}
\usepackage{rotating}

\begin{document}

\title{
Dosimetry characterization of $^{32}$P intravascular brachytherapy
source wires using Monte Carlo codes PENELOPE and GEANT4}

\author{Javier Torres and Manuel J. Buades}

\affiliation{Servicio de Radiof\'{\i}sica y Protecci\'on
Radiol\'ogica, Hospital Universitario ``Virgen de la Arrixaca'',
E-30120 Murcia, Spain.}

\author{Julio F. Almansa and Rafael Guerrero}

\affiliation{Servicio de Radiof\'{\i}sica y Protecci\'on Radiol\'ogica, 
Hospital Universitario ``Puerta del Mar'',
 E-11009 C\'adiz, Spain.}

\author{Antonio M. Lallena}

\affiliation{Departamento de F\'{\i}sica Moderna, 
Universidad de Granada, E-18071 Granada, Spain.}

\vspace*{1cm}

\begin{abstract}
Monte Carlo calculations using the codes PENELOPE and GEANT4 have been
performed to characterize the dosimetric parameters of the new 20~mm
long catheter based $^{32}$P beta source manufactured by Guidant
Corporation. The dose distribution along the transverse axis and the
two dimensional dose rate table have been calculated. Also, the dose
rate at the reference point, the radial dose function and the
anisotropy function were evaluated according to the adapted TG-60
formalism for cylindrical sources. PENELOPE and GEANT4 codes were
first verified against previous results corresponding to the old 27~mm
Guidant $^{32}$P beta source. The dose rate at the reference point for
the unsheathed 27~mm source in water was calculated to be $0.215 \pm
0.001$~cGy~s$^{-1}$~mCi$^{-1}$, for PENELOPE, and $0.2312 \pm
0.0008$~cGy~s$^{-1}$~mCi$^{-1}$, for GEANT4. For the unsheathed 20~mm
source these values were $0.2908 \pm 0.0009$~cGy~s$^{-1}$~mCi$^{-1}$
and $0.311 \pm 0.001$~cGy~s$^{-1}$~mCi$^{-1}$, respectively. Also, a
comparison with the limited data available on this new source is
shown. We found non negligible differences between the results
obtained with PENELOPE and GEANT4.
\end{abstract}


\maketitle

\section{Introduction}

One of the major limitations of coronary interventions lies in the
appearance of restenosis of the treated vessel. Restenosis occurs in
35\%-40\% of the treated patients in case of angioplasty, whilst the
percentage reduces to 22\%-32\% in case coronary stents are used
\cite{Poc95}.

Intravascular brachytherapy appears to be a useful tool to solve such
problems \cite{Fox02}. PREVENT \cite{Rai99} and INHIBIT \cite{Wak00}
studies have pointed out the ability of ionizing radiation to inhibit
restenosis. Though either gamma and beta radiation have been used,
endovascular beta-radiotherapy has become most popular in Europe due
to the obvious radiation safety advantages.

This new, rapidly growing interest has led both the AAPM \cite{Nat99}
and the ESTRO \cite{Pot01} to report recommendations on the medical
physic issues. At the minimum standard level, dose at the
prescription point in the central plane is encouraged to be recorded
and reported. Additional dose calculations in peripheral planes
require complete two-dimensional (2D) dosimetry information.  However,
beta sources present a high dose gradient, the distances in which the
dose is deposited being considerably small. This has made it difficult
to perform adequate dosimetry on intravascular sources.

Recently, the dosimetric characterization of a 27 mm long, catheter
based, $^{32}$P beta source wire, manufactured by Guidant Corporation
(Houston, Texas, USA), has been done by means of plastic scintillator
\cite{Mou00} and radiochromic film \cite{Mou00,Boh01}.

Monte Carlo (MC) simulations have proved to be a valuable complementary
means for determining the dose distributions around beta
emitters. The $^{32}$P source mentioned above has been simulated with
different MC codes (MCNPv4B2 \cite{Mou00}-\cite{Seh01},
CYLTRAN/ITS3 \cite{Mou00,Sel02}, MCNPv4C \cite{Mou00}, EGS4
\cite{Tod00,Wan01,Li01} and EGSnrc \cite{Wan01}) and the calculated
doses agree within $<10$\% with the experimental results.

In this work we address the 2D dosimetric characterization of a new,
clinically used, 20 mm long catheter based $^{32}$P beta source
wire, commercially available in the Galileo$^{\rm TM}$ Intravascular
Radiotherapy System, manufactured also by Guidant Corporation. This
source has been used \cite{Mou03} in a detailed study in which a
variety of MC codes are intercompared. Also a comparison with
measurements done by using radiochromic-dye film and an extrapolation
chamber is carried out. Measured and calculated values were found to
agree within 10\% in a polystyrene phantom. On the other hand, the
calculated depth dose curves, using the MC codes, were within 5\% at
4~mm depth, after an adequate selection of the tracking parameters. In
this last case water was considered the medium surrounding the source.

Here we have used the MC codes PENELOPE \cite{Sal01} and GEANT4
\cite{Geant}. PENELOPE was designed specifically to account for the
transport of low-energy electrons, photons and positrons and has been
used to simulate other brachytherapy sources \cite{San98,Ase02}.
GEANT4 includes special packages which permit to describe low energy
physics processes and, at present, it is being used to simulate
different situations of medical physics interest \cite{Geant}. In this
work we have determined the dose rate at the reference point, the
depth dose and the two dimensional dose distribution. We also
calculated the radial dose and the anisotropy functions, as proposed
in the adapted AAPM TG-60 formalism for cylindrical sources
\cite{Sch01}, used also in Ref. \cite{Wan01}. We compare some of our
results with those quoted in Ref. \cite{Sel02}.

\begin{figure}
\begin{center}
\epsfig{file=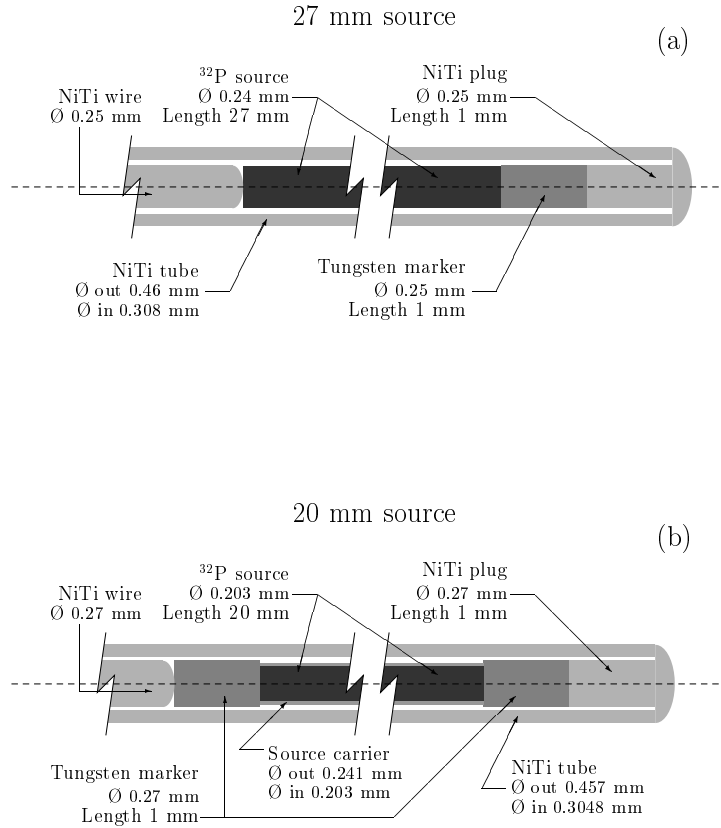,width=10cm}
\caption{\small 
Schematic view of the 27 (upper panel) and 20~mm
(lower panel) $^{32}$P sources manufactured by Guidant Corporation.}
\end{center}
\end{figure}

\section{Material and Methods}

\subsection{Source models}

A schematic view of the geometry of the two sources analyzed is shown
in Fig. 1. The 27~mm source has been described in
Refs. \cite{Mou00,Boh01}. The geometry of the 20~mm source has been
obtained from Ref. \cite{Sel02} and corresponds closely to that quoted
in \cite{Mou03}. The active cores are cylinders with length of 27 and
20~mm, respectively. These cores are made of C$_2$H$_6$O$_2$.  The
20~mm source presents a 0.019~mm thick source carrier wrapping the
active core.  This is made of polymide. The cores are encapsulated by
NiTi tubes with thickness of 0.076~mm in both sources.  In the 27~mm
source, a tungsten X-ray marker of 1~mm in length follows the core at
the distal end. In the case of the 20~mm source, two markers of the
same characteristics are located at both the distal and the proximal
ends of the active core. Both sources include a NiTi plug of 1~mm in
length following the distal tungsten X-ray marker. A NiTi
hemispherical seal is welded at the distal end and an unbound NiTi
wire is attached to the proximal end in both sources.  Empty regions
between the different pieces and the capsules are supposed to be
filled with air.

Sources are centered inside a catheter of Nylon-12. In some of the
calculations done for the 27~mm source, we have considered a
helicoidal catheter (labeled catheter 1) reported in
Refs. \cite{Mou00}-\cite{Wan01}. In the case of the 20~mm source, we
have calculated using both this helicoidal catheter and a lobed
catheter (labeled catheter 2) currently in use. This catheter is a new
design to increase the blood flow during the procedure. Both catheters
were modeled as tubes with an external radius of 0.4445~mm, while the
inner diameter was 0.597~mm for catheter 1 and 0.5716~mm for catheter
2.  Gaps between catheters and sources are assumed to be filled with
air.

Catheter and source were placed in a water phantom the dimensions of
which are large enough to ensure no significant loss of radiation.

The composition and density considered in the simulation for the different
materials is given in Table \ref{tab:compos}.

\begin{table}[hb]
\begin{tabular}{ccccccc}
\hline \hline
& ~~C$_2$H$_6$O$_2$~~ & ~~NiTi~~ & ~~Tungsten~~ 
& ~~Polymide~~~ & Air & ~~~Nylon-12~~ \\
\hline\hline
H  & 0.097432 &      &              & 0.026362 &          & 0.11749  \\
C  & 0.387026 &      &              & 0.691133 & 0.000124 & 0.73045  \\
N  &          &      &              & 0.073270 & 0.755267 & 0.07098  \\
O  & 0.515542 &      &              & 0.209235 & 0.231781 & 0.08108  \\
Ar &          &      &              &          & 0.012827 &          \\
Ti &          & 0.44 &              &          &          &          \\
Ni &          & 0.56 &              &          &          &          \\
W  &          &      &  1.0         &          &          &          \\
\hline
density [g~cm$^{-3}$] & 1.1155 & 6.5 & 19.3 & 1.096 & 0.0012048 & 1.196 \\
\hline \hline
\end{tabular}
\caption{\small
Composition of the different materials used in the MC simulations
performed in this work. The values correspond to the weight fraction
of each element in the material. In the case of C$_2$H$_6$O$_2$, the
density quoted is that used in the 20~mm source. For the 27~mm source
a density of 1.12~g~cm$^{-3}$ has been used.
\label{tab:compos}}
\end{table}

\subsection{Monte Carlo calculations}

In this work the PENELOPE \cite{Sal01} and GEANT4 \cite{Geant} Monte
Carlo codes have been used to perform the simulations. PENELOPE is a
general purpose MC code which allows to simulate the coupled
electron-photon transport. It can be applied to energies ranging from
a few hundred eV up to 1~GeV and with arbitrary materials. Besides,
PENELOPE permits a good description of the particle transport at
interfaces and presents an accurate description of the electron
transport at low energies. These characteristics make PENELOPE a
useful tool for medical physics applications as previous works have
pointed out (see e.g. Refs. \cite{San98,Ase02}). We have used the
version 2001 of the code. Details about the physical processes
included can be found in Ref. \cite{Sal01}.

GEANT4 is a toolkit aiming to simulate the passage of particles
through matter for applications in high energy physics, nuclear
experiments, medical physics studies, etc. Due to its purpose, the
available energy range and the possible particles to be simulated are
much larger than for PENELOPE. GEANT4 includes low energy packages
which extend the energy range of particle interactions for electrons,
positrons and photons down to 250 eV, and can be used up to
approximately 100 GeV. In this work we have used the version 4.1 with
the patch-01 and the low energy package G4EMLOW1.1. 
Detailed documentation concerning the physics involved can
be found in Ref. \cite{Geant}.

As the sources considered here present cylindrical symmetry, a
cylindrical coordinate system was used with the origin in the center
of the active core and with the $z$ axis along the source and pointing
to its distal end. The dose distributions were supposed to be
functions of the axial coordinate $z$ and the radial coordinate $\rho$.
The position of the reference point is
($\rho_0=2$~mm, $z_0=0$).

The radionuclide was assumed to be uniformly distributed inside the
active core. The direction of the emitted electrons was supposed to be
isotropic around the initial position. The initial electron energies
were sampled from the corresponding Fermi distributions, taken from
Ref. \cite{Gar85}. A calculated average energy of $0.6957$~MeV was
obtained, in agreement with values reported in previous works
\cite{Boh01,Vyn82}.

In PENELOPE analog simulation is performed for photons. For electrons,
instead, the simulation is done in a mixed scheme in which collisions
characterized by a scattering angle larger than a certain value are
called hard collisions and are individually simulated, while
collisions with a scattering angle smaller that the limit value are
called soft collisions and are described by means of a multiple
scattering theory. The electron tracking is controlled by means of
four parameters. $C_1$ and $C_2$ refer to elastic collisions. $C_1$
gives the average angular deflection due to a elastic hard collision
and to the soft collisions previous to it. $C_2$ represents the
maximum value permitted for the average fractional energy loss in a
step. On the other hand, $W_{\rm cc}$ and $W_{\rm cr}$ are energy
cutoffs to distinguish hard and soft events. Thus, the inelastic
electron collisions with energy loss $W<W_{\rm cc}$ and the emission
of Bremsstrahlung photons with energy $W<W_{\rm cr}$ are considered in
the simulation as soft interactions. In our calculations the
simulation parameters were fixed to the following values: $W_{\rm
cc}=5$~keV, $W_{\rm cr}=1$~keV, $C_1=C_2=0.05$, $s_{\rm
max}=10^{35}$. This last parameter is an upper bound for the length of
the steps generated in the simulation.

Photons were simulated down to 1~keV. Primary
electrons, $\delta$ rays and positrons were absorbed when they slow
down to kinetic energies of 1~keV, in air, and 50~keV, in all the
remaining materials, except in the active core where the cut off
energy was fixed to 200~keV. This value takes care of the fact that
electrons with such an energy are not expected to leave the active
core. Those betas generated with energies below the simulation
threshold were not transported and their energy was deposited
at the source position.

In case of GEANT4 simulations, it is necessary to choose a given range
threshold for each particle type. This threshold is converted,
internally, to an energy threshold below which secondary particles are
not emitted. In our calculations the range thresholds have been fixed
to 100~nm providing energy thresholds of 1 keV for both photons and
electrons in all materials considered.

The statistical uncertainties were calculated by scoring both the
energy deposited in each voxel and its square for each history. These
voxels were taken to be annular bins with thicknesses $\Delta
\rho=0.1$~mm and $\Delta z=0.2$~mm. The average energy deposited in a
given voxel (per incident particle) is
\[
 E_{\rm mean} \, = \, \frac{1}{N} \sum_{i=1}^{N} e_{i} \, ,
\]
where $N$ is the number of simulated histories and $e_{i}$ is the
energy deposited by all the particles of the $i$-th history (that is,
including the primary particle and all the secondaries it
generates). The statistical uncertainty is given by
\[
 \sigma_{E_{\rm mean}} \, = \, 
 \sqrt{\frac{1}{N} \left[ \frac{1}{N}\sum_{i=1}^{N} e_{i}^2 
\, - \, {E_{\rm mean}}^2 \right] } \, .
\]
The uncertainties given throughout the paper correspond to
1$\sigma$. In tables, and in order to simplify the writing, they are
given between parentheses; for example, a value quoted as 0.956(2)
means $0.956 \pm 0.002$.

In case of the 27~mm source, the simulations have been done by
following $2\times 10^7$ histories for PENELOPE and $5\times 10^7$ for
GEANT4. For the 20~mm source a total of $5\times 10^7$ histories were
simulated with both codes. These numbers permitted to keep the
statistical uncertainties under reasonable levels.

Additionally, we performed, for the 20~mm source, simulations
with PENELOPE by using scoring voxels given by cylindrical annuli of
$\Delta z=13$~mm centered along the source axial direction, and
$\Delta \rho=0.1$~mm in the radial direction. A total of $6\times
10^7$ histories were followed in this case. These scoring voxels
warranty radial distributions which are effectively constant over the
axial length and permit the uncertainty to be reduced considerably.

\section{Results}

\subsection{27~mm source}

First we validated the MC codes with the old 27~mm source. Our results
were compared with those reported in literature. 

In Refs. \cite{Mou00} and \cite{Boh01} the MCNPv4B2 code was
used. Mourtada {\it et al.} \cite{Mou00} transported $10^6$ histories
in each run, with scoring voxels given by cylindrical annuli of 20~mm
in the source axial direction and 0.1~mm in the radial
direction. They showed also the dose rate per unit activity along the
transverse axis. The same code and voxels were used by Bohm {\it et al.}
\cite{Boh01} for the dose rate per unit activity along the transverse
axis. Besides, these authors calculated the dose to water, away and
along the source, using thin cylindrical shells with a thickness of
0.1~mm and a height of 0.1~mm as scoring voxels.  Wang and Li
\cite{Wan01} used scoring voxels equal to those we have considered
here. Finally, Seltzer \cite{Sel02} quoted the results obtained
for a simulation with CYLTRAN/ITS3, but did not give information about
the scoring voxels.

The dose rate at the reference point for the unsheathed 27~mm source
in water was calculated to be $0.215 \pm 0.001$ and $0.2312 \pm
0.0008$~cGy~s$^{-1}$~mCi$^{-1}$ for PENELOPE and GEANT4,
respectively, thus differing by 7\%. These values must be
compared with the results obtained with the other codes. The
calculation with EGSnrc \cite{Wan01} provides $0.2185 \pm
0.0002$~cGy~s$^{-1}$~mCi$^{-1}$. Mourtada {\it et al.}  \cite{Mou00}
obtained 0.229~cGy~s$^{-1}$~mCi$^{-1}$ by means of MCNPv4B2. Finally,
in Ref. \cite{Boh01} a value of $0.232 \pm
0.00028$~cGy~s$^{-1}$~mCi$^{-1}$ was found also with MCNPv4B2. This is
the largest value found, which differs from the one we have obtained
with PENELOPE (the smallest one) by $\sim 8$\%.

\begin{table}[ht]
\begin{tabular}{ccccccccc}
\hline \hline
$z$ & $\rho$ & \multicolumn{2}{c}{this work} & Ref. \protect\cite{Boh01} 
& Ref. \protect\cite{Sel02}
& \multicolumn{2}{c}{Ref. \protect\cite{Wan01}} \\ \cline{3-4} \cline{7-8}
{[mm]} & {[mm]} & PENELOPE & GEANT4
       & MCNPv4B2  & CYLTRAN/ITS3 & EGSnrc & EGS4 \\  \hline\hline
 0.0 & 1.0   & 0.718(3)  & 0.721(2)  & 0.728  & 0.801 & 0.731   & 0.729   \\  
     & 2.0   & 0.221(1)  & 0.2379(9) & 0.232  & 0.232 & 0.225   & 0.222   \\  
     & 3.0   & 0.0718(6) & 0.0868(4) & 0.0785 & 0.0767 & 0.0732  & 0.0708  \\ 
     & 4.0   & 0.0202(3) & 0.0281(2) & 0.0229 & 0.0212 & 0.0207  & 0.0193  \\ \hline
10.0 & 1.0   & 0.719(3)  & 0.717(2)  & 0.725   & & &  \\  
     & 2.0   & 0.220(1) & 0.2356(9) & 0.230    & &  &  \\  
     & 3.0   & 0.0708(6) & 0.0853(4) & 0.0769  & &  &  \\ 
     & 4.0   & 0.0206(3) & 0.0275(2) & 0.0223  & &  &  \\ \cline{1-5}
12.0 & 1.0   & 0.677(3)  & 0.667(2)& 0.681     & & &  \\  
     & 2.0   & 0.193(1) & 0.2046(8) & 0.202    & &  &  \\  
     & 3.0   & 0.0609(6) & 0.0713(4) &0.0654   & &  &  \\ 
     & 4.0   & 0.0167(3) & 0.0230(2) & 0.0186  & &  &  \\ \cline{1-5}
14.0 & 1.0   & 0.180(2)  & 0.187(1) & 0.188    & & &  \\  
     & 2.0   & 0.0755(8) & 0.0816(5) & 0.0793  & &  &  \\  
     & 3.0   & 0.0266(4) & 0.0322(3) & 0.0291  & &  &  \\ 
     & 4.0   & 0.0076(2) & 0.0109(1) & 0.00870 & &  &  \\ \cline{1-5}
16.0 & 1.0   & 0.0097(5) & 0.0130(3) & 0.0113  & & &  \\  
     & 2.0   & 0.0067(3) & 0.0101(2) & 0.00858 & &  &  \\  
     & 3.0   & 0.0035(1) & 0.0051(1) & 0.00423 & &  &  \\ 
     & 4.0   & 0.00111(7) & 0.00192(6) & 0.00135 & &  &  \\
\hline \hline
\end{tabular}
\caption{\small
Two-dimensional dose rates (in cGy~s$^{-1}$~mCi$^{-1}$) for the 27~mm
source sheathed by catheter 1. The errors corresponding to the
calculations performed with MCNPv4B2 (Ref. \protect\cite{Boh01}) as
well as those with EGSnrc and EGS4 (Ref. \protect\cite{Wan01}) are
quoted to be below 1\%. The CYLTRAN/ITS3 data have been obtained by
scanning Fig. 11 of Ref. \protect\cite{Sel02} and therefore are
indicative only.
\label{tab:td-27C}}
\end{table}

\begin{table}[ht]
\begin{tabular}{ccccccc}
\hline \hline
$\rho$ && \multicolumn{5}{c}{$z$ [mm]} \\ \cline{3-7}
 [mm] & MC code & 0.0 & 10.0 & 12.0 & 14.0 & 16.0 \\
\hline\hline
 & EGSnrc  & 3.2844 & 3.2730 & 3.0795 & 0.8230 & 0.0439 \\
1.0 
 & PENELOPE & 3.31(1)  & 3.29(1)  &  
  3.11(1) & 0.835(8) & 0.062(2) \\
 & GEANT4   & 3.036(9) & 3.023(9) &  
  2.835(8) & 0.789(5) & 0.051(1) \\ 
\hline
 & EGSnrc  & 1.0000 & 0.9926 & 0.8762 & 0.3363 & 0.0324 \\
2.0 
 & PENELOPE & 1.000(6) & 0.996(6) &  
  0.881(6) & 0.336(3) & 0.039(1) \\
 & GEANT4   & 1.000(4) & 0.988(4) &  
  0.860(3) & 0.346(2) & 0.0394(8) \\ 
\hline
 & EGSnrc  & 0.3226 & 0.3194 & 0.2711 & 0.1176 & 0.0158 \\
3.0 
 & PENELOPE & 0.323(3) & 0.317(3) &  
  0.271(3) & 0.117(2) & 0.0182(7) \\
 & GEANT4   & 0.358(2) & 0.352(2) &  
  0.299(2) & 0.131(1) & 0.0209(5) \\ 
\hline
 & EGSnrc  & 0.0897 & 0.0890 & 0.0744 & 0.0336 & 0.0050 \\
4.0 
 & PENELOPE & 0.089(1) & 0.091(1) &  
  0.074(1) & 0.0329(8) & 0.0050(3) \\
 & GEANT4   & 0.1153(9) & 0.1125(9) &  
  0.0957(8) & 0.0453(6) & 0.0079(2) \\
\hline \hline
\end{tabular}
\caption{\small
Two-dimensional dose rates for the unsheathed 27~mm source. The values
are normalized to the dose rate at the reference point ($\rho_0=2$~mm,
$z_0=0$). The errors corresponding to the calculations performed with
EGSnrc (Ref. \protect\cite{Wan01}) are quoted to be below 1\%.
\label{tab:td-27}}
\end{table}

In Table \ref{tab:td-27C} we compare the dose rate per unit activity
along the transverse axis (values for $z=0$ in the first set of rows)
obtained with PENELOPE and GEANT4 with those of
Refs. \cite{Boh01,Wan01,Sel02}. The results correspond to the $^{32}$P
wire source sheathed by catheter 1. As we can see PENELOPE produces
results very similar to those of EGSnrc. In this case the differences
for any $\rho$ value are below 6\% with respect to the dose rate
obtained with PENELOPE at the reference point, $\dot{D}_{\rm
PEN}(\rho_0,z_0)$. Up to $\rho=3$~mm PENELOPE practically coincides
with EGS4, the differences relative to $\dot{D}_{\rm PEN}(\rho_0,z_0)$
being smaller than 5\%. The difference between GEANT4
(MCNPv4B2) and PENELOPE is below 8\% (5\%),
relative to $\dot{D}_{\rm PEN}(\rho_0,z_0)$. In what respect to
the results of CYLTRAN/ITS3, they show negligible differences with
PENELOPE results. The fact that these last quoted values have been
obtained by scanning Figure 11 of Ref. \cite{Sel02} makes these
differences to be only indicative.

Two-dimensional dose distributions are shown in Tables
\ref{tab:td-27C} and \ref{tab:td-27} for some values of $\rho$ and
$z\geq 0$. Similar results are obtained for the negative values of
$z$, though the source wire is not symmetric with respect to $z=0$. In
Table \ref{tab:td-27C} our results are compared with those of Bohm
{\it et al.} \cite{Boh01} calculated with MCNPv4B2 for the source
sheathed by catheter 1. In Table \ref{tab:td-27} the dose
distributions for the unsheathed source, normalized to the reference
point, are compared with the results obtained by Wang and Li
\cite{Wan01} using EGSnrc.

The differences between MCNPv4B2 and PENELOPE (see Table
\ref{tab:td-27C}), remain below 5\%, relative to $\dot{D}_{\rm
PEN}(\rho_0,z_0)$). Same situation appears for MCNPv4B2 and GEANT4,
which give very similar results for $\rho \leq 2$~mm.

Table \ref{tab:td-27} shows that EGSnrc and PENELOPE provide very
similar normalized doses. It is important to remember here that the
difference in the doses at the reference point between both codes is
1.6\% only. The differences maintain at more or less the same level
for all $z$ and $\rho$ values studied.

Finally, and as we can see in both tables, the discrepancies between
PENELOPE and GEANT4, observed for the dose rate per unit activity
along the transverse axis, show up again. In any case the differences
remain below 8\% with respect to $\dot{D}_{\rm PEN}(\rho_0,z_0)$.

\subsection{20~mm source}

\subsubsection{Unsheathed source}

Once we have analyzed the results obtained for the 27~mm source, we
will discuss the dose rates found for the 20~mm source. All the
results here correspond to the unsheathed source. For clarity, we
leave the analysis of the sheathed source for the next subsection.

\begin{table}
\begin{rotate}{-90}
\begin{tabular}{ccccccccccc}
\hline \hline
$z$ & \multicolumn{10}{c}{$\rho$ [mm]} \\ \cline{2-11} 
[mm] & 0.4 & 0.6 & 0.8 & 1.0 & 1.5 & 2.0 & 2.5 & 3.0 & 4.0 & 5.0 \\ 
\hline
     & \multicolumn{3}{c}{$\Delta z=13$~mm} & 0.9576(3)  & 0.5139(2)
 & 0.2918(1)  & 0.16689(9) & 0.09386(6) & 0.02618(3) &            \\ \cline{2-10} 
 0.0 & 2.771(6)   & 1.788(4)   & 1.279(3)   & 0.961(2)   & 0.512(1)   
 & 0.2908(9)  & 0.1667(6)  & 0.0936(4)  & 0.0263(2)  & 0.00552(9) \\
     & 2.544(6)   & 1.730(4)   & 1.256(3)   & 0.952(2)   & 0.527(1)   
 & 0.311(1)   & 0.1877(7)  & 0.1133(5)  & 0.0364(3)  & 0.0091(1) \\\hline
 2.0 & 2.770(6)   & 1.790(4)   & 1.275(3)   & 0.956(2)   & 0.517(1)   
 & 0.2912(9)  & 0.1670(6)  & 0.0940(4)  & 0.0260(2)  & 0.00561(9) \\
     & 2.542(6)   & 1.736(4)   & 1.259(3)   & 0.955(2)   & 0.526(1)   
 & 0.314(1)   & 0.1875(7)  & 0.1123(5)  & 0.0365(3)  & 0.0092(1) \\\hline
 4.0 & 2.771(6)   & 1.790(4)   & 1.274(3)   & 0.957(2)   & 0.514(1)   
 & 0.2899(9)  & 0.1673(6)  & 0.0937(4)  & 0.0263(2)  & 0.00533(8) \\
     & 2.545(6)   & 1.727(4)   & 1.256(3)   & 0.947(2)   & 0.524(1)   
 & 0.310(1)   & 0.1873(7)  & 0.1129(5)  & 0.0362(3)  & 0.0091(1) \\\hline
 6.0 & 2.765(6)   & 1.786(4)   & 1.272(3)   & 0.952(2)   & 0.511(1)   
 & 0.2895(9)  & 0.1670(6)  & 0.0923(4)  & 0.0261(2)  & 0.00552(9) \\
     & 2.549(6)   & 1.727(4)   & 1.258(3)   & 0.950(2)   & 0.525(1)   
 & 0.312(1)   & 0.1866(7)  & 0.1124(5)  & 0.0361(3)  & 0.0090(1) \\\hline
 7.0 & 2.761(6)   & 1.788(4)   & 1.267(3)   & 0.947(2)   & 0.507(1)   
 & 0.2866(9)  & 0.1635(6)  & 0.0920(4)  & 0.0253(2)  & 0.00532(8) \\
     & 2.533(6)   & 1.719(4)   & 1.244(3)   & 0.938(2)   & 0.516(1)   
 & 0.305(1)   & 0.1819(7)  & 0.1092(5)  & 0.0347(2)  & 0.0087(1) \\\hline
 8.0 & 2.746(6)   & 1.761(4)   & 1.243(3)   & 0.929(2)   & 0.487(1)   
 & 0.2729(9)  & 0.1525(6)  & 0.0853(4)  & 0.0233(2)  & 0.00479(8) \\
     & 2.507(6)   & 1.698(4)   & 1.223(3)   & 0.912(2)   & 0.495(1)   
 & 0.2870(9)  & 0.1712(7)  & 0.1008(5)  & 0.0319(2)  & 0.0079(1) \\\hline
 9.0 & 2.650(6)   & 1.663(4)   & 1.152(3)   & 0.839(2)   & 0.422(1)   
 & 0.2305(8)  & 0.1262(6)  & 0.0702(4)  & 0.0191(2)  & 0.00397(7) \\
    & 2.410(6)   & 1.598(4)   & 1.122(3)   & 0.825(2)   & 0.427(1)   
 & 0.2421(9)  & 0.1407(6)  & 0.0827(4)  & 0.0261(2)  & 0.0066(1) \\\hline
10.0 & 1.389(4)   & 0.891(3)   & 0.635(2)   & 0.471(2)   & 0.254(1)   
 & 0.1444(7)  & 0.0823(4)  & 0.0465(3)  & 0.0131(2)  & 0.00268(6) \\
    & 1.271(4)   & 0.857(3)   & 0.620(2)   & 0.470(2)   & 0.261(1)   
 & 0.1541(7)  & 0.0940(5)  & 0.0557(3)  & 0.0181(2)  & 0.00449(8) \\\hline
11.0 & 0.086(1)   & 0.100(1)   & 0.1070(9)  & 0.1064(8)  & 0.0841(6)  
 & 0.0588(4)  & 0.0378(3)  & 0.0231(2)  & 0.0068(1)  & 0.00139(4) \\
     & 0.094(1)   & 0.110(1)   & 0.114(1)   & 0.1132(9)  & 0.0924(6)  
 & 0.0668(5)  & 0.0457(3)  & 0.0291(3)  & 0.0100(1)  & 0.00247(6) \\
\hline \hline
\end{tabular}
\end{rotate}
\vspace*{16cm}
\caption{\small
Two-dimensional dose rates (in cGy~s$^{-1}$~mCi$^{-1}$) for
the unsheathed 20~mm source in water, calculated with PENELOPE (first
row) and GEANT4 (second row) for each $z$ value.  All the results have
been obtained using voxels with $\Delta z=0.2$~mm, except those of the
first row corresponding to $z=0$ which have been calculated with
PENELOPE using voxels with $\Delta z=13$~mm.
\label{tab:td-dr-P}}
\end{table}

In Table \ref{tab:td-dr-P} (box labeled $z=0$) we show the results we
have obtained for the dose rate per unit activity along the transverse
axis. The first row gives the results obtained using PENELOPE for
voxels with $\Delta z=13$~mm. The second (third) rows correspond to
the results obtained with PENELOPE (GEANT4) for voxels with $\Delta
z=0.2$~mm. The dose rates at the reference point calculated with
PENELOPE are $0.2918 \pm 0.0001$~cGy~s$^{-1}$~mCi$^{-1}$ when the
large scoring voxel (that with $\Delta z=13$~mm) is used and $0.2908
\pm 0.0009$~cGy~s$^{-1}$~mCi$^{-1}$ for the small one (that with
$\Delta z=0.2$~mm). GEANT4 produces a value of $0.311 \pm
0.001$~cGy~s$^{-1}$~mCi$^{-1}$, which differs from the previous ones
in $\sim 7$\%. On the other hand, the discrepancy between the results
obtained with both codes increases with increasing $\rho$, showing a
behavior similar to that found for the 27~mm source.

\begin{figure}[hb]
\begin{center}
\epsfig{file=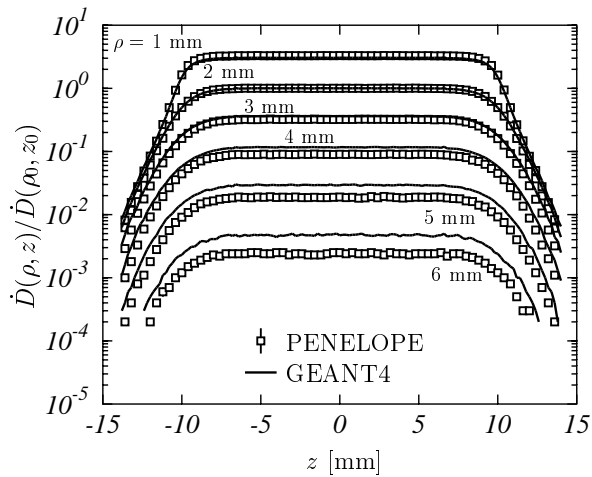,width=8cm}
\caption{\small
Axial dose distributions for the unsheathed 20~mm
source in water, for $\rho=1$ to 6 mm. The values are normalized to
the reference point ($\rho_0=2$~mm, $z_0=0$). PENELOPE results are
plotted with points, while curves correspond to GEANT4.}
\end{center}
\end{figure}

\begin{figure}[ht]
\begin{center}
\epsfig{file=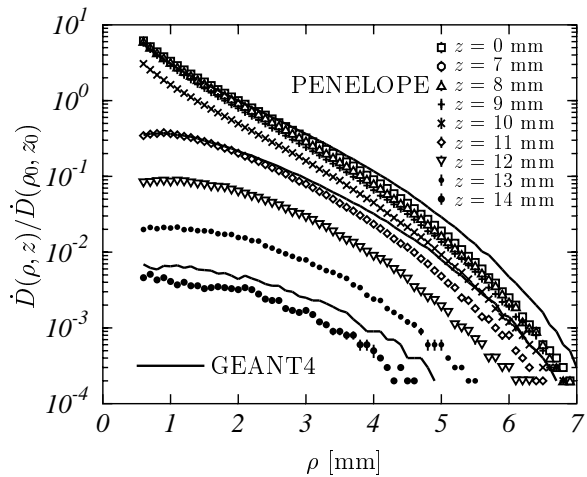,width=8cm}
\caption{\small
Radial dose distributions for the unsheathed 20~mm
source in water, for $z=0$ and $z=7$ to 14~mm. The values are
normalized to the reference point ($\rho_0=2$~mm, $z_0=0$). PENELOPE
results are plotted with points.  GEANT4 results corresponding to
$z=0$, 11 and 14~mm are plotted with lines.}
\end{center}
\end{figure}

Fig. 2 shows the comparison of the axial dose
distributions (normalized to the corresponding reference points)
obtained with both codes for $\rho$ values ranging from 1 to 6 mm.
The disagreement between PENELOPE and GEANT4 shows up also in
Fig. 3. Therein the dose distributions along transverse axes
(normalized to the reference point) for $z$ values ranging from 7 to
14 mm are compared to those corresponding to $z=0$ mm. PENELOPE
results are plotted with symbols, while GEANT4 calculations are only
shown (curves) for $z=0$, 11 and 14 mm. For a given $\rho$ value the
discrepancy grows with $z$. In general, GEANT4 predicts larger doses at
large $\rho$ values. The case of the $z=11$ mm curve is a clear
example of this behavior: up to $\rho \sim 3$ mm both results are in
a reasonable agreement but for $\rho \geq 5$ mm the GEANT4 result
coincides with the PENELOPE one obtained for $z=10$ mm.

In the case of the 27~mm source, Wang and Li \cite{Wan01} quoted an
asymmetry of the dose distribution due to the presence of the distal
tungsten X-ray marker. Such asymmetry showed up for $z>13$~mm and the
dose rates at the proximal and distal ends showed a relative
difference below 25\%. If this is so, the presence of two such
markers, in the 20~mm source, should make this asymmetry to
disappear. Fig. 4 shows the ratio of the dose at various $z$ values to
the dose at $-z$. The results are similar in both codes. As we can
see, the ratio of the dose rates does not show statistically
significant differences from unity, even at large $z$. This means that
the asymmetry of the source structure due to the small differences
between the distal and the proximal ends is smaller than statistical
uncertainties and would therefore probably be irrelevant in practice.

\begin{figure}
\begin{center}
\epsfig{file=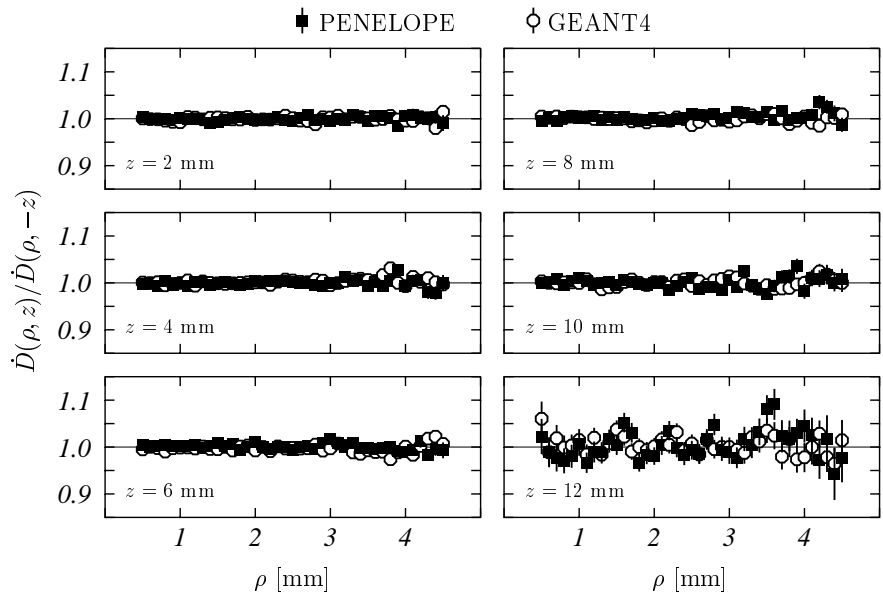,width=12cm}
\caption{\small
Ratio of the dose at different $z$ values to the dose
at $-z$ as a function of $\rho$, for the unsheathed 20~mm source in
water. PENELOPE (black squares) and GEANT4 (open circles) results are
shown.}
\end{center}
\end{figure}

\begin{figure}
\begin{center}
\epsfig{file=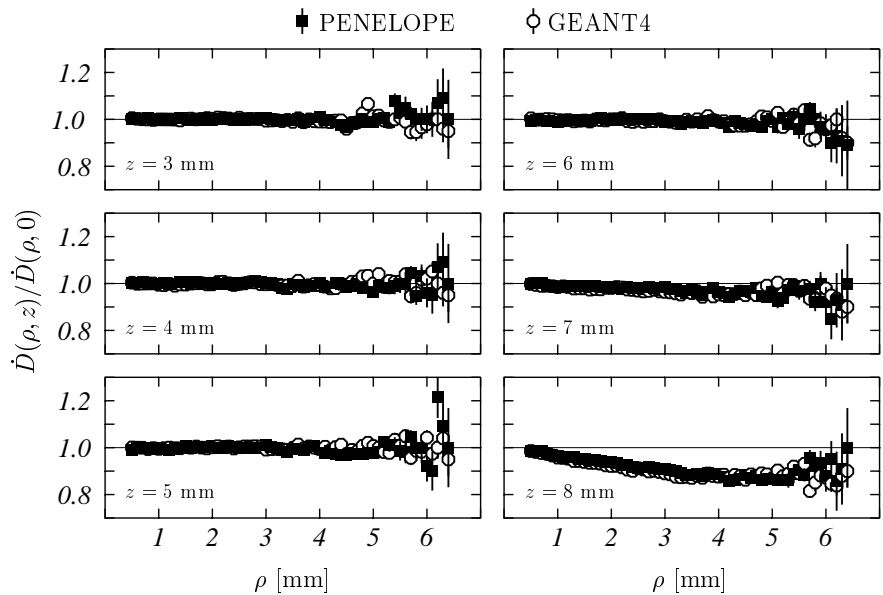,width=12cm}
\caption{\small
Ratio of the dose at different $z$ values to the dose
at $z=0$ as a function of $\rho$, for the unsheathed 20~mm source in
water. Results obtained with PENELOPE (GEANT4) are shown with black
squares (open circles).}
\end{center}
\end{figure}

\begin{table}[hb]
\begin{tabular}{cccccc}
\hline \hline
$\rho$ & \multicolumn{2}{c}{20~mm source} &~& 
\multicolumn{2}{c}{27~mm source} \\
\cline{2-3} \cline{5-6} 
[mm] & PENELOPE & GEANT4 & &PENELOPE & EGSnrc \\ \hline
 0.5 & 1.698(7)  & 1.507(6)  &
& 1.75(1)   & 1.7410\\
 1.0 & 1.542(7)  & 1.428(6)  &
& 1.57(1)   & 1.5582\\
 1.5 & 1.275(6)  & 1.226(6)  &
& 1.30(1)   & 1.2867\\
 2.0 & 1.000(5)  & 1.000(5)  &
& 1.000(9)  & 1.0000\\
 2.5 & 0.742(4)  & 0.780(4) &
& 0.730(7)  & 0.7343\\
 3.0 & 0.518(3)  & 0.586(3)  &
& 0.511(5)  & 0.5093\\
 3.5 & 0.342(2)  & 0.412(3)  &
& 0.334(4)  & 0.3310\\
 4.0 & 0.208(2)  & 0.270(2)  &
& 0.199(3)  & 0.1990\\
 4.5 & 0.119(1)  & 0.167(2)  &
& 0.108(2)  & 0.1098\\
 5.0 & 0.059(1)  & 0.090 (1)  &
& 0.054(2)  & 0.0543\\
 5.5 & 0.0257(6) & 0.0451(9) &
& 0.025(1)  & 0.0231\\
 6.0 & 0.0099(4) & 0.0190(6) &
& 0.0093(7) & 0.0083\\
 6.5 & 0.0028(2) & 0.0070(3) &
& 0.0034(4) & 0.0025\\
 7.0 & 0.0006(1) & 0.0017(2) &
& 0.0008(2) & 0.0007 \\ 
\hline \hline
\end{tabular}
\caption{\small
Radial dose functions for the $^{32}$P sources. 
Simulations were performed in water.
\label{tab:rdf}}
\end{table}

In Fig. 5 we analyze the axial dose uniformity by
plotting the ratio of the dose at $z=3$ up to 8 mm to the dose at
$z=0$ as a function of $\rho$. The results show that, for a given
$\rho$ value, doses stay constant for axial positions up to 6-7 mm. This
occurs for both PENELOPE and GEANT4 and means that the scoring voxels
with $\Delta z=13$~mm, considered above to reduce the uncertainty in
calculating the radial dose, can be used safely for that
purpose. 

The two-dimensional dose rates we have obtained using voxels with
$\Delta z=0.2$~mm are listed in Table \ref{tab:td-dr-P}. For each $z$
value, the first row corresponds to PENELOPE and the second one to
GEANT4.  The values for $z<0$ are not included because of the symmetry
with respect to the corresponding $z>0$ results. On the other hand,
the uniformity up to $z \sim 6$ mm is again evident. The comparison
between both codes indicates that GEANT4 predicts larger doses
than PENELOPE except for $\rho$ below 0.5 mm.

To finish with the analysis of the unsheathed source, we calculated the
radial dose functions and the anisotropy functions. Following
Ref. \cite{Sch01}, these quantities are given in the cylindrical
coordinate system instead of the spherical one used in the TG-60
formalism \cite{Nat99}. For the radial dose we have
\begin{equation}
g(\rho) \, = \, \displaystyle 
\frac{\dot{D}(\rho,z_0)}{\dot{D}(\rho_0,z_0)} \, 
\frac{G(\rho_0,z_0)}{G(\rho,z_0)} \, .
\end{equation}
The anisotropy function is given by
\begin{equation}
F(\rho,z) \, = \, \displaystyle
\frac{\dot{D}(\rho,z)}{\dot{D}(\rho,z_0)} \, 
\frac{G(\rho,z_0)}{G(\rho,z)} \, ,
\end{equation}
where $G(\rho,z)=\beta/(L \rho)$, with $L=27$ or 20~mm, according to
the source considered, and $\beta$ is the angle between the lines
joining the point ($\rho$,$z$) and the two extremes of $L$.

Calculated radial dose functions for PENELOPE and GEANT4 in water are
shown in Table \ref{tab:rdf}. As we can see, the discrepancies
observed between both codes in previous results are evident here
again. Besides, we have included the results obtained with PENELOPE
for the 27~mm source and those quoted by Wang and Li for EGSnrc
\cite{Wan01}. The radial dose function for the 20~mm source appears to
be rather similar to that of the 27~mm source. The results obtained
with PENELOPE differs less than 10\% up to $\rho =6$~mm. Finally,
PENELOPE and EGSnrc provide results which coincide within the
statistical uncertainties for the 27~mm source.

\begin{figure}
\begin{center}
\epsfig{file=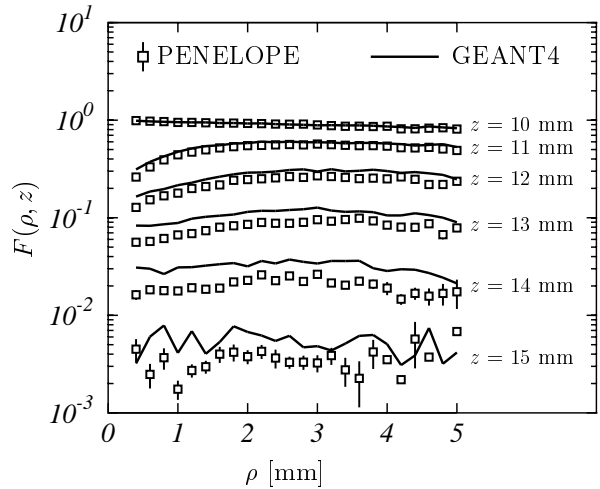,width=8cm}
\caption{\small
Comparison between the anisotropy functions for the
unsheathed 20~mm source obtained with PENELOPE (squares) and GEANT4
(curves).  The errors for GEANT4 results are of the same order than
those obtained with PENELOPE.}
\end{center}
\end{figure}

Table \ref{tab:any-20-P} shows the anisotropy function calculated for
the unsheathed 20~mm source with PENELOPE. These results are compared
in Fig. 6 with those obtained with GEANT4. As in previous comparisons,
the discrepancies between both codes appear to be noticeable for
large $\rho$ and/or $z$ values, though the absolute differences
remain small in comparison with the values obtained at the reference
point.

\begin{table}[hb]
\begin{tabular}{ccccccccccc}
\hline \hline
$z$ & \multicolumn{10}{c}{$\rho$ [mm]} \\ \cline{2-11} 
[mm] & 0.4 & 0.6 & 0.8 & 1.0 & 1.5 & 2.0 & 2.5 & 3.0 
& 4.0 & 5.0 \\ \hline
   0.0 & 1.000(3)   & 1.000(3)   & 1.000(3)   & 1.000(4)   & 1.000(4)   
& 1.000(5)   & 1.000(6)   & 1.000(7)   & 1.00(1)    & 1.00(2) \\
   2.0 & 1.001(3)   & 1.003(3)   & 0.999(3)   & 0.998(4)   & 1.014(4)   
& 1.007(5)   & 1.009(6)   & 1.012(7)   & 1.00(1)    & 1.03(2) \\
   4.0 & 1.005(3)   & 1.009(3)   & 1.006(3)   & 1.009(4)   & 1.024(4)   
& 1.022(5)   & 1.035(6)   & 1.038(7)   & 1.05(1)    & 1.02(2) \\
   6.0 & 1.013(3)   & 1.022(3)   & 1.025(4)   & 1.028(4)   & 1.055(4)   
& 1.071(5)   & 1.094(6)   & 1.091(8)   & 1.12(1)    & 1.14(3) \\
   7.0 & 1.022(3)   & 1.039(3)   & 1.042(4)   & 1.050(4)   & 1.087(4)   
& 1.111(5)   & 1.130(6)   & 1.153(8)   & 1.16(1)    & 1.17(3) \\
   8.0 & 1.038(3)   & 1.057(3)   & 1.067(4)   & 1.084(4)   & 1.122(5)   
& 1.148(5)   & 1.151(7)   & 1.169(8)   & 1.16(1)    & 1.14(3) \\
   9.0 & 1.068(3)   & 1.094(4)   & 1.108(4)   & 1.115(4)   & 1.128(5)   
& 1.128(6)   & 1.102(7)   & 1.106(8)   & 1.08(1)    & 1.05(3) \\
  10.0 & 0.990(4)   & 0.977(4)   & 0.967(4)   & 0.949(4)   & 0.944(5)   
& 0.927(5)   & 0.905(6)   & 0.894(7)   & 0.87(1)    & 0.81(2) \\
  11.0 & 0.261(4)   & 0.331(4)   & 0.392(4)   & 0.441(4)   & 0.512(4)   
& 0.548(5)   & 0.561(5)   & 0.569(6)   & 0.542(9)   & 0.49(2) \\
  12.0 & 0.128(4)   & 0.153(3)   & 0.168(3)   & 0.179(3)   & 0.218(3)   
& 0.246(4)   & 0.260(4)   & 0.269(4)   & 0.250(6)   & 0.24(1) \\
  13.0 & 0.056(3)   & 0.057(2)   & 0.061(2)   & 0.067(2)   & 0.077(2)   
& 0.086(2)   & 0.089(2)   & 0.095(3)   & 0.084(4)   & 0.079(5) \\
  14.0 & 0.016(2)   & 0.018(2)   & 0.018(1)   & 0.018(1)   & 0.018(1)   
& 0.023(1)   & 0.023(1)   & 0.026(2)   & 0.019(3)   & 0.017(6) \\
  15.0 & 0.005(1)   & 0.0025(7)  & 0.0037(9)  & 0.0018(4)  & 0.0034(7)  
& 0.0038(6)  & 0.0031(4)  & 0.0032(6)  & 0.00351(3) & 0.0068(1) \\
\hline \hline
\end{tabular}
\caption{\small
Anisotropy function $F(\rho,z)$ for
the unsheathed 20~mm source in water, calculated with PENELOPE.
\label{tab:any-20-P}}
\end{table}

\subsubsection{Sheathed source}

Here we analyze the effect of the catheter on the dosimetry. First we
compare the results we have obtained with both codes with the scarce
data available in the literature. Table \ref{tab:dd-20-c1} shows this
comparison with the data provided by the manufacturer in the source
manual (second column) and with the results of a set of calculations
with different codes performed by Seltzer \cite{Sel02}. In these
calculations the scoring voxels were taken to be annuli of radial
thickness 0.1 mm and axial length 15 mm, centered along the active
wire length. This produces greatly reduced statistical
uncertainties. As can be seen, our results for PENELOPE coincide with
those found with CYLTRAN/ITS3 and PENELOPE in Ref. \cite{Sel02},
except for small deviations (less than 1\%) at $\rho = 2$~mm in the
last case. The values provided by Guidant are between 5 and 10\% above
those we have found for PENELOPE. MCNP4C results are similar to the
Guidant ones and those calculated with EGS4 and EGSnrc fall to $\sim
7$\% below those obtained with PENELOPE at $\rho=4$~mm. The values
found with GEANT4 follow the same trend seen in the previous sections.

\begin{table}[ht]
\begin{tabular}{cccccccccc}
\hline \hline
$\rho$ & & 
\multicolumn{5}{c}{Ref. \protect\cite{Sel02}} & ~ &
\multicolumn{2}{c}{this work} \\ \cline{3-7} \cline{9-10}
{[mm]} & Guidant & 
CYLTRAN/ITS3 & MCNP4C & EGS4 & EGSnrc & PENELOPE & 
& PENELOPE & GEANT4 \\  \hline
1.0 & 0.999  & & & & & & 
& 0.975(2)  & 0.971 (2) \\  
1.5 & 0.548  & & & & & &
& 0.526(1)  & 0.540 (1) \\
2.0 & 0.317  & 0.300  & 0.314  & 0.296  & 0.296  & 0.298  &  
& 0.3005(9) & 0.321 (1) \\  
2.5 & 0.184  & & & & & &
& 0.1723(6) & 0.1957(7) \\ 
3.0 & 0.105  & & & & & &
& 0.0976(4) & 0.1172(5) \\ 
3.5 & 0.0569 & & & & & &
& 0.0534(3) & 0.0683(4) \\ 
4.0 & 0.0290 & 0.0276 & 0.0285 & 0.0255 & 0.0254 & 0.0274 &
& 0.0274(2) & 0.0384(3) \\ 
4.5 & 0.0135 & & & & & &
& 0.0133(1) & 0.0200(2) \\
\hline \hline
\end{tabular}
\caption{\small
Depth doses (in cGy~s$^{-1}$~mCi$^{-1}$) for the 20~mm source
sheathed by catheter 1. The values of the second column are those provided
by the manufacturer in the GALILEO$^{\rm TM}$ system manual.
Simulations were performed in water. 
\label{tab:dd-20-c1}}
\end{table}

\begin{figure}
\begin{center}
\epsfig{file=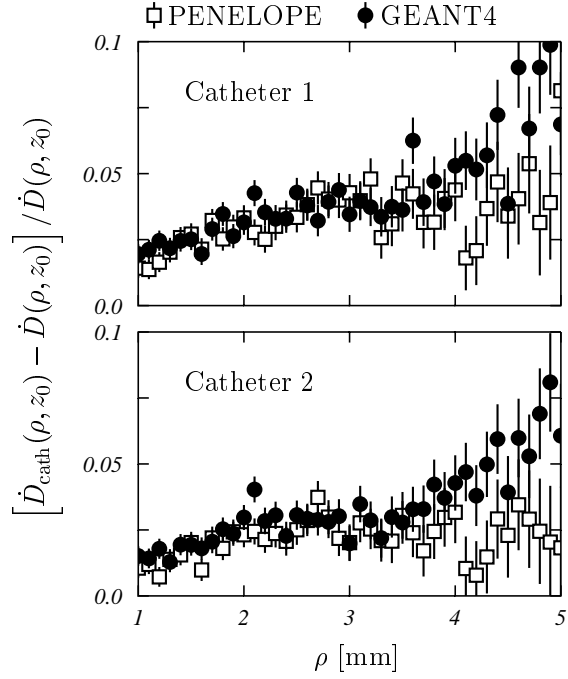,width=8cm}
\caption{\small
 Relative difference between the depth dose rates
calculated for the 20~mm source sheathed, $\dot{D}_{\rm
cath}(\rho,z_0)$, and unsheathed, $\dot{D}(\rho,z_0)$, with 
PENELOPE (squares) and GEANT4 (disks). Upper (lower) panel
corresponds to catheter 1 (2).}
\end{center}
\end{figure}

Fig. 7 shows the relative difference
\begin{equation}
\displaystyle
\frac{\dot{D}_{\rm cath}(\rho,z_0)-\dot{D}(\rho,z_0)}
{\dot{D}(\rho,z_0)}
\end{equation}
between the depth doses calculated with and without catheter for
PENELOPE (squares) and GEANT4 (disks). The results corresponding to
the 20~mm source are shown. As we can see, both codes produce similar
results with both catheters. However, GEANT4 shows an increase of the
effect of the catheter above $\rho=3.5$~mm which is not present in the
PENELOPE calculation. Similar results are obtained for the 27~mm
source sheathed by catheter 1. On the other hand, catheter 2 produces
a modification of the dose in water slightly smaller than catheter
1. This last gives rise to enhancements of the dose rate on the
transverse axis reaching  $\sim 4$\% above $\rho=3$~mm, whilst the
first one increases the dose rate by  $\sim 2.5$\% at most. This
result is reasonable if one considers that catheter 2 has a smaller
inner radius than catheter 1, thus reducing the air gap. In any case,
uncertainties in source strength measurements might outweigh this
correction in clinical practice.

\section{Conclusions}

In this work we have performed the dosimetric characterization of two
wire sources of $^{32}$P (of 27~mm and 20~mm active length)
manufactured by Guidant Corporation and clinically used in
intravascular brachytherapy. To do so we have used the Monte
Carlo codes PENELOPE and GEANT4.

First we have validated the codes by comparing the results obtained
with these two codes with those available in the literature for the
27~mm source. The main conclusion is that PENELOPE produces dose rates
in reasonable agreement with those quoted in Ref. \cite{Wan01} for
EGSnrc and EGS4 except in the tails of the axial doses (large $z$
values). Calculations performed with MCNPv4B2 \cite{Mou00,Boh01}
differ from those obtained with PENELOPE as the distance to the
source increases. The comparison between PENELOPE and GEANT4 shows a
clear disagreement. 

For the 20~mm source, we have calculated the depth doses on the
transverse axis, the two-dimensional dose rate table and the radial
dose and anisotropy functions (this two last according to the approach
proposed in Ref. \cite{Sch01} for cylindrical sources). We have
checked that the dose distribution remains uniform along the axial
direction for $|z|$ smaller than 6-7~mm and $\rho \sim 7$~mm. Besides,
we have tested the symmetry of the dose distribution. The dose rates
calculated with PENELOPE show a good agreement with the scarce data
available, while those obtained with GEANT4 follow a trend similar to
that observed in the 27~mm source.

The presence of the catheter sheathing the sources produces an
increase in the dose rate on the transverse axis. This enhancement is
larger for the helicoidal catheter (catheter 1) than for the lobed one
(catheter 2). For the first one, the dose on the transverse axis
increases up to 4\% above $\rho=3$~mm, while the second one produces a
maximum enhancement of $\sim 2.5$\%. These differences could produce
non-negligible modifications in the clinical use of the sources.

The results we have obtained indicate that PENELOPE produces a dose
rate map which can be considered to represent the dosimetric
characterization of the sources here studied. The appreciable
discrepancies shown by GEANT4 results need a deeper investigation
in order to clarify the origin of these differences. In particular, to
determine if they are due to the physics included in the simulations
or to the tracking procedure should be of great interest. We are
working in this direction.

The full data set we have obtained here is available at
http://fm131.ugr.es/P32-IVB

\acknowledgments{
Authors wish to acknowledge Guidant Corporation (European division)
for the assistance provided concerning the geometrical details of the
sources here investigated. We are also indebted to Dr. Jenkins
(Guidant, Texas) for his patient answering to our questions.  This
work has been supported in part by the Junta de Andaluc\'{\i}a
(FQM0225).
}

\end{document}